# Transcriptional profiling reveals fundamental differences in iPS-derived progenitors of endothelial cells (PECs) versus adult circulating EPCs

**Elmira Jalilian, William Raimes**

[1] Institute of Ophthalmology, University College London, London EC1V 9EL, United Kingdom

*Corresponding author: jalilian@med.umich.edu

**Keywords:** iPSC-derived progenitors of endothelial cells, cord-derived EPCs; peripheral-blood EPCs, Stem cell; CD34; Gene transcriptional profiling; EC differentiation

**Abstract**

There are a number of different stem cell sources that have the potential to be used as therapeutics in vascular degenerative diseases. On the one hand, there are so called endothelial progenitor cells (EPCs), which are typically derived from adult blood. They carry the marker CD34, but the true nature and definition of EPCs is still controversial. On the other hand, there are embryonic precursors of endothelial cells (PECs), which also express CD34, and can be differentiated from embryonic stem cells (ESCs) or induced pluripotent stem cells (iPSCs) in vitro. In this study, it was aimed to compare these two different CD34 positive cell populations by full genome transcriptional profiling (RNAseq). To this end, we firstly optimised a PEC differentiation protocol and found that vascular endothelial growth factor (VEGF) is critical for the transition of cells from mesodermal precursors to PECs. Additionally, principal component analysis (PCA) of RNAseq data showed that blood-derived EPCs clustered far from iPS-derived PECs which illustrates these populations are fundamentally different. This data will be useful to better define these cell populations and facilitating the translation of regenerative approaches in this field as well as providing potentially novel diagnostic tools.

**Introduction**

Progenitors of endothelial cells are known to be a promising stem cell source for vascular regeneration [1-4]. These cells were initially shown in human adult circulating blood cells[5] and were called "endothelial progenitor cells" (EPC). These cells were later on shown to be present in adult bone marrow (BM)[6-8] and peripheral blood (PB)[9]. EPCs were believed to be progenitor cells and could differentiate into endothelial cells in vitro. However, the controversy over the origin, differentiation and cellular identity of these cells remains a potential issue[10]. CD34 is the main marker of these cells[11,12]. Despite a high number of studies for the clinical usefulness of CD34+ cells[13-15], it is still not completely understood what exactly these cells are and what are the exact role of this population outside of each individual specificity[16]. On one hand, a small population of circulating cells in adult peripheral blood stream has been shown to generate CD34+ cells. It has been suggested that these cells can be further stimulated to migrate, proliferate or differentiate into a more mature lineage and are able to either directly [17-19] contribute or indirectly [20-22] support, vascular regeneration. On the other hand, during embryonic development, endothelial cells develop from a precursor's population[23]. These "true" embryonic progenitor cells also express CD34, which are which are also shown to participate in vessel regeneration[24]. These embryonic-derived progenitors (PECs) are most likely to be very different from adult peripheral blood CD34+ cells. To date it is not well understood what these differences are. Therefore, comparative studies of different CD34+ cells populations based on their broad range of molecular characteristics may contribute to understanding the difference between all these populations and would offer an insight into overlapping properties of the cells that express CD34. In this study, we intended to use gene expression profiling to better characterise the identity of circulating adult EPCs and iPS-derived PECs. Thus, we first aim to establish an efficient protocol to generate iPS-derived CD34+ cells in sufficient quantity in vitro and then compare their gene expression profiling with Adult CD34+ cells isolated from cord blood and peripheral blood.

**Results and Discussion**

*In vitro derivation of progenitors of endothelial cells (PECs) from from induced pluripotent stem cells (iPSCs)*

In order to establish a protocol to generate high yield of PECs in vitro, sequential experiments were developed. Key parameters such as factors, concentrations and time course of factors and substrates were modified in further continuous experiments to optimise the best conditions in which to grow PECs. Each experiment provided continuous feedback which was used for the further optimization process. At the end of experimental process cells were fixed with 4% PFA and then processed for immune-staining to assess the CD34 expression as a PEC marker. Outcome from the first set of experiments were used to change the design for the next experiment. Conditions with more cell death and no CD34 expression were excluded. Better conditions were selected to repeat in the next experiments in addition to new conditions. In all experiments each treatment condition was performed in triplicate. Relative expression of CD34 was analysed in a semi-quantitative score method explained in method 1.6. Briefly, relative expression of CD34 was assessed as relative CD34+ cell yield (number sign), cell detachment (low/ medium/ high) and CD34+ coverage (disperse/ aggregated / extensive) for each treatment condition.

First step of optimization was to find the most suitable medium. Six different media were tested with different combinations of factors in 96 well plates including the media used in two papers [25,26]. Among all the media tested, DMEM/F12 in combination with B2+ N27 was the best medium to induce the most CD34 positive cells and it was selected for the remainder of the experiments. Data shown in (**Supplementary Fig. 1**). Over the optimization procedures the concentrations of Activin A 25 ng/ml, VEGF 50 ng/ml and bFGF 25ng/ml were not changed. BIO and SB431512 were used in the concentration of 1.5 μM and 10 μM initially. However, conditions having BIO 1.5 μM had a big area of cell death presumably because of the high concentration of BIO (**Supplementary Fig. 2**). Therefore, the concentration of BIO was further reduced to 0.5 μM and finally adjusted at final concentration of 0.15 μM. Furthermore, conditions containing SB 431542 10 μM from day 2, also showed high amount of cell death, Therefore, in further experiments, the concentration of SB 431542 was reduced to 2 μM (**Supplementary Fig**.**3**). After optimisation of medium and concentrations of factors, since enough CD34 enrichment was not observed an attempt was made to optimise the protocols by modifying different parameters such as substrate and different combinations of factors. Furthermore, since variable results were observed over repeated experiments up to this point, it was assumed that the number of cells initially plated might also be an effective factor in the induction of CD34+ cells. Therefore, in the next experiment substrate was modified in parallel with three different cell numbers (low, medium and high) to find out its influence to enhance iPS-derived CD34+ cells.

Furthermore, to get advantage of 3D culture system and create a more physiological microenvironment, extra Matrigel was added on top of the medium and then compared with 2D culture condition. iPSC cells were passaged and plated in high density (about 40,000 cells), middle density (25000 cells) and low density (13000 cells) in 96 well plates coated with Matrigel. Four different conditions were tested. to create 3D culture condition additional 2.5% Matrigel (1:80 final working ratios), was mixed into the medium and added to the plates and similar conditions were compared to 2D culture conditions (conditions summarised in Error! Reference source not found.). Immunostaining results showed that adding extra Matrigel (2.5%) on top of the cells could dramatically increase the CD34 yield in all conditions (Error! Reference source not found.).

Furthermore, the highest expression of CD34 was observed in middle and high density. Conditions 1 and 2 had the highest expression of CD34 in middle and high density in 3D culture system whereas conditions 3 and 4 had the highest expression in low and middle density (Error! Reference source not found.). Therefore, it was concluded seeding cells in high density (40000 cells/well) and 3D culture condition is sufficient to induce the most CD34 positive cells. Moreover, consistent results using bFGF in the treatments were not observed in repeated experiments. Since CD34 expression was observed in conditions without bFGF, this factor was excluded from later experiments.

| Treatment no. | | Experiment 4 (iPSC, DM+Extra 2.5% Matrigel) | | | | Relative CD34+ yield | Cell detachment | CD34+ coverage |
|---|---|---|---|---|---|---|---|---|
| | | Day 1 | Day 2 | Day 4 | Day 6 | | | |
| Low density, with 2.5% Matrigel | 1 | ActA/B4/V/BIO** | ActA/B4/V/BIO** | V/S | Fixed | | Medium | Dispersed |
| | 2 | ActA/B4/V/BIO** | B4/V/BIO** | V/S | Fixed | | Medium | Aggregated |
| | 3 | ActA/B4/V/F | F/B4/V | F/B4/V | Fixed | | Low | Aggregated |
| | 4 | B4 | F/V | F/V | Fixed | | Low | Aggregated |
| Low density, without Matrigel | 1 | ActA/B4/V/BIO** | ActA/B4/V/BIO** | V/S | Fixed | | Low | Dispersed |
| | 2 | ActA/B4/V/BIO** | B4/V/BIO** | V/S | Fixed | | Low | N/A |
| | 3 | ActA/B4/V/F | F/B4/V | F/B4/V | Fixed | | Medium | N/A |
| | 4 | B4 | F/V | F/V | Fixed | | Medium | N/A |
| Middle density, with 2.5% Matrigel | 1 | ActA/B4/V/BIO** | ActA/B4/V/BIO** | V/S | Fixed | | Low | Extensive |
| | 2 | ActA/B4/V/BIO** | B4/V/BIO** | V/S | Fixed | | Low | Extensive |
| | 3 | ActA/B4/V/F | F/B4/V | F/B4/V | Fixed | | Low | Aggregated |
| | 4 | B4 | F/V | F/V | Fixed | | Low | N/A |
| Middle density, | 1 | ActA/B4/V/BIO** | ActA/B4/V/BIO** | V/S | Fixed | | Low | Aggregated |
| | 2 | ActA/B4/V/BIO** | B4/V/BIO** | V/S | Fixed | | Medium | Aggregated |
| without Matrigel | 3 | ActA/B4/V/F | F/B4/V | F/B4/V | Fixed | | Medium | N/A |
| | 4 | B4 | F/V | F/V | Fixed | | Medium | N/A |
| High density, with 2.5% Matrigel | 1 | ActA/B4/V/BIO** | ActA/B4/V/BIO** | V/S | Fixed | | Low | Extensive |
| | 2 | ActA/B4/V/BIO** | B4/V/BIO** | V/S | Fixed | | Low | Extensive |
| | 3 | ActA/B4/V/F | F/B4/V | F/B4/V | Fixed | | Low | Aggregated |
| | 4 | B4 | F/V | F/V | Fixed | | Low | N/A |
| High density, without Matrigel | 1 | ActA/B4/V/BIO** | ActA/B4/V/BIO** | V/S | Fixed | | Low | Aggregated |
| | 2 | ActA/B4/V/BIO** | B4/V/BIO** | V/S | Fixed | | Low | Aggregated |
| | 3 | ActA/B4/V/F | F/B4/V | F/B4/V | Fixed | | Medium | N/A |
| | 4 | B4 | F/V | F/V | Fixed | | Medium | N/A |

| -/# | ## | ###/#### | ##### | ####### |
|---|---|---|---|---|

**Table 1.** Different treatment conditions in experiment 4. Cells were seeded onto Matrigel-coated 96-well plates at $1.3 \times 10^4$ (low density), $2.5 \times 10^4$ (medium density) and $4 \times 10^4$ (high density) cells/well. Four different conditions were used. 3D culture conditions (2.5% Matrigel was added on top of the medium) were compared to 2D culture conditions (no Matrigel on top). Experimental timelines ran 6 days long and consisted of growth factors addition in fresh medium on day 1, 2 and 4. Cells were fixed and stained on day 6. Immunocytochemistry data were analysed in a semi-quantitative score method mentioned previously. iPSC: induced pluripotent stem cell; ActA: Activin A 25 ng/ml; B4: Bone Morphogenetic Protein 4, 30 ng/ml; V: Vascular Endothelial Growth Factor (VEGF) 50 ng/ml; F: basic Fibroblast Growth Factor 25 ng/ml; BIO**: 6-Bromoindirubin-3'-oxime (Wnt pathway activator) 0.15 μM; S: SB431542 (inhibitor of ALK5, ALK4 and ALK7) 2 μM.

In further optimisation process, the effect of BMP4 and BIO were assessed in different time points and in different combinations. (**Supplementary Fig**. **4**)[27-29]. Immunocytochemistry results showed that BMP4 for three days is more effective than one day in inducing CD34 expression. Furthermore, conditions containing BMP4 from day 1 to 5 alone or in combination with other factors did not show any expression of CD34 which could indicate that BMP4 is a critical signalling molecule required for inducing the mesodermal lineage only in the very early phase of mesodermal induction. However, having BMP4 alone is not effective and BMP4 should be applied in combination with other factors.

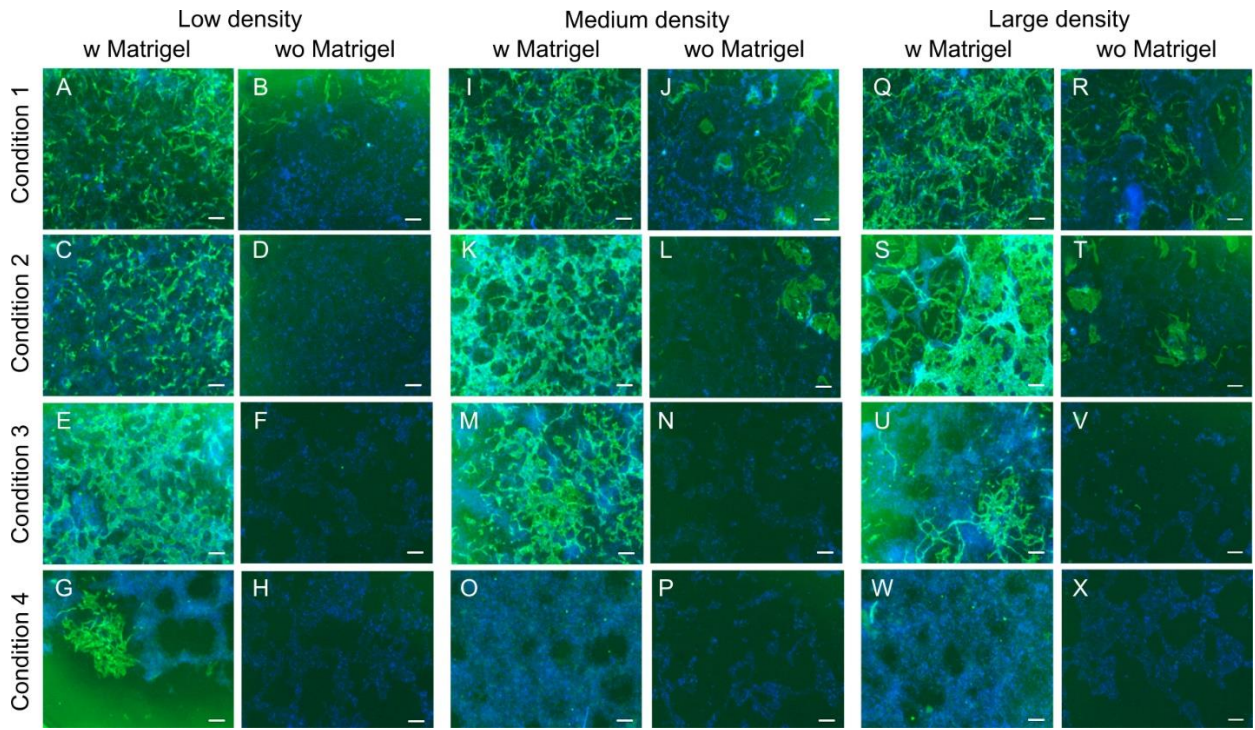

**Figure 1** Immunofluorescent staining of the expression of PECs marker CD34 (green). Four treatment conditions were used to compare 3D versus 2D environments in three different cell densities (low/medium/high). Conditions having extra Matrigel on top of the medium as a multi-layer environment had remarkable expression of CD34 (A, C, E, I, K, M, Q, S & U) compared to 2D culture conditions (B, D, F, J, L, N, R, T and V) in all treatment conditions except condition 4 that did not show any considerable expression between two different culture systems (G, O, W vs. H, P, X). Furthermore, conditions 1 and 2 showed increasing trend in CD34 expression from low density to high density in both 3D and 2D culture systems whereas conditions 4 and 5 had decreasing trend in CD34 expression from high density to low density in both 3D and 2D culture systems (E, F, M, N, U, V and G, H, Q, P, W, X). Nuclei are stained with DAPI in blue. Scale bar, 164 μm.

Next, the effect of BMP4 alone, condition 10 in (**Supplementary Fig. 4**) or in combination with different factors was assessed for different time points and it was compared with the condition without BMP4. It was intended to find out if BMP4 would be sufficient enough to induce the high amount of CD34 or adding other factors to the cocktail is important to induce the CD34 expression. Condition with only VEGF treatment was also considered, condition 11 in (**Supplementary Fig. 4**) to assess if VEGF alone would be enough to induce CD34 expression. Previous experiments could clearly illustrate that low concentration of SB 431542 (2μM) had much lower cell death compared to high concentrations (10μM) which was again confirmed in current experiment. In current experiment SB 431542 (2μM) was also used at different time points to find out its effect in increasing the CD34 expression and comparing this to conditions without SB 431542. Furthermore, the effect of BIO for the first 1 or 3 days was also assessed in combination with other factors, all conditions are shown in (**Supplementary Fig. 4**). Similar to previous experiments, cells were seeded in triplicate onto Matrigel-coated 96-well plates in 3D culture system at $4\times10^4$ cells/well. Experimental timelines ran 6 days long and consisted of growth factors addition in fresh medium on day 1, 2 and 4. Cells were fixed and stained on day 6.  were fixed and stained on day 6.

Immunocytochemistry results showed that BMP4 for three days is more effective than one day in inducing CD34 expression. However, having BMP4 alone, condition 10 (**Supplementary Fig. 4**) is not effective and BMP4 should be applied in combination with specific factors. From current experiment, it could be concluded that BIO is a critical factor to be considered with BMP4 in the cocktail. Conditions containing BMP4 but not having BIO had no or very low expression of CD34 (conditions 1, 2, 3, 10 and 11). The presence of BIO either for one or three days seems to have high influence in inducing the CD34 positive cells (conditions 5, 6 and 7). However, it is not clear how would be the influence of BIO if it would be kept longer up to day 5. Therefore, this hypothesis was assessed in the next experiment. Furthermore, SB 431542 was another important factor to be considered albeit in combination with specific factors and specific time period. Conditions without SB 431542 had very low or no expression of CD34, conditions 1, 2, 10 and 11 (**Supplementary Fig. 4**) which could show the importance of this factor in CD34 induction. Moreover, conditions with the highest expression of CD34 had SB 431542 only for two days. Conditions containing SB 431542 for more than 2 days (from day 2 to 6) had a big area of cell death and no or very low expression of CD34, conditions 3, 4, 8 and 9 (**Supplementary Fig. 4**). This observation could suggest that SB 431542 only for the last two days is sufficient enough to induce the most CD34 positive cells. This was consistent with previous findings showing that inhibition of TGF-ß before mesodermal induction results in a reduction of CD34 expression[30]. Therefore, SB 431542 was considered to apply for only two days during the late phase of treatment (day three to five). Moreover, conditions containing BMP4 from day 1 to 5 alone, conditions 10 (**Supplementary Fig. 4**) or in combination with other factors, condition 1 (**Supplementary Fig. 4**) did not show any expression of CD34 which could indicate that BMP4 is a critical signaling molecule required for inducing the mesodermal lineage only in the very early phase of mesodermal induction. Additionally, having VEGF (50ng/ml) alone was not enough to induce the CD34 expression, condition 11 (**Supplementary Fig. 4**). This experiment could show the importance of BIO and BMP4 in CD34 induction. Since this experiment could show the importance of BIO in CD34 induction, next experiment was designed to further investigate the longer effect of BIO and its effect in combination with or without BMP4 and SB 431542, all data shown in (**Supplementary Fig. 4**).

From the fact that applying BIO either for 1 or 3 days was shown to be very effective in inducing the high number of CD34, further experiment was designed to find out the longer effect of BIO factor in the absence or presentence of SB 431542 or BMP4 in CD34 induction. BIO was considered for all conditions for either 1, 2 or 5 days. Furthermore, BMP4 was also considered for either 1 or 3 days which was combined with SB 431542 in some of the conditions to compare the different combinations of factors. Furthermore, some conditions evaluated the effect of BIO in absence of BMP4 with or without SB 431542 (all conditions shown in supplementary data 5). Immunocytochemistry results revealed that conditions adding BIO to the medium for longer time period (day zero to five) suppressed the expression of CD34 positive cells, condition 3 and 4 in (**Supplementary Fig. 5**).

BIO is necessary only for the early phase of treatment but becomes detrimental at later stages. Furthermore, the combination of this factor with SB 431542 (condition 3 supplementary data 5) from day 3 to 5 had relatively bigger area of cell detachment compared to condition without SB 431542. Conditions having BIO and BMP4 for 3 days and then VEGF + SB 431542 for 2 days had comparatively more CD34 expression compared to similar condition without SB 431542, Condition 5 vs. 6 (**Supplementary Fig. 5**). Furthermore, it was observed that conditions having BIO for 3 days but excluding BMP4 from day 2 conditions 5 & 6 (**Supplementary Fig. 5**) still had high expression of CD34 and it was comparatively more if SB 431542 was added to the cocktail in the last 2 days. However, if BIO excluded from day 2 and instead BMP4 gets included up to day 3 condition 7 and 8 (**Supplementary Fig. 5**), less CD34 was observed compared to 3 days stimulation with BIO. Similarly, having SB 431542 had slightly higher CD34 coverage compared to VEGF only condition in the last 2 days of treatment. In condition 9 and 10 BMP4 is excluded from day 1 and is only included in condition 10 from day 2 to 5. Comparison of the two condition shows that including BMP4 is effective in increasing the number of CD34 positive cells. Comparisons of conditions 9 and 11 supplementary data 5 which are similar and the only difference is the presence or absence of SB 431542 emphasize the positive effect of SB 431542 in inducing of the CD34 positive cells. Furthermore, excluding all factors except VEGF from day 2 does not seem to be effective in CD34 induction (condition 12). Condition without any of these factors does not show any CD34 expression condition 13. Comparing all possible factor compositions shows that including BIO and BMP4 together for three days is more effective than having each of them separately for 3 days or including one of them for 1 day and the other for 3 days. Furthermore, including SB 431542 for the last two days is effective for increasing the number of CD34 positive cell. This was consistent with previous findings showing that inhibition of TGF-ß before mesodermal induction results in a reduction of CD34 expression [30]. Moreover, in all current experiment Activin A was removed from day 2. Comparing immune-staining results from this experiment with previous experiment which contained Activin A for three days reveals that presence of Activin A for three days is more effective than 1 day. Therefore, up to this point, combining Activin A, BIO, BMP4 and VEGF for the first day and then SB 431542 and VEGF only for two days seems to be very effective to induce the greatest number of CD34 positive cells (**Supplementary. Fig**. **5**).

*VEGF is critical for transition of cells from mesodermal lineage into progenitor of endothelial cells*
VEGF signalling is shown to be crucial for vessel development and patterning and also for endothelial cell sprouting and migration [31-33]. Since VEGF-A has been identified as an essential factor for endothelial differentiation, it was initially used from day zero till to day-5. Comparing different conditions with or without VEGF over different experiments showed that VEGF is the key factor for generation of CD34+ cells and no expression was detected without VEGF. So far, we and others used VEGF right from the beginning for all differentiation protocols. However, up to now, no one has clearly shown in which specific stage of differentiation VEGF is crucial to generate progenitors of endothelial cells. Therefore, in current study, we intended to find out if VEGF is needed from day zero up to day five. Or if excluded from different time points during the treatment, whether it would still affect the induction of endothelial progenitor cells treatment conditions are shown in table in (Error! Reference source not found.). This experiment was repeated three times and similar results were observed form both experiments. (n=3 biological replicate and n=9 technical replicate). We found that, adding VEGF only during the late phase of treatment and for only two days was sufficient to induce the expression of CD34 positive cells. Interestingly, Immunocytochemistry results of cells just before adding VEGF at day three, showed faint CD34 expression, arrows in NO VEGF condition in (Error! Reference source not found.). These weakly CD34 positive cells might be "precursors of progenitors of endothelial cells", here designated "PPECs".

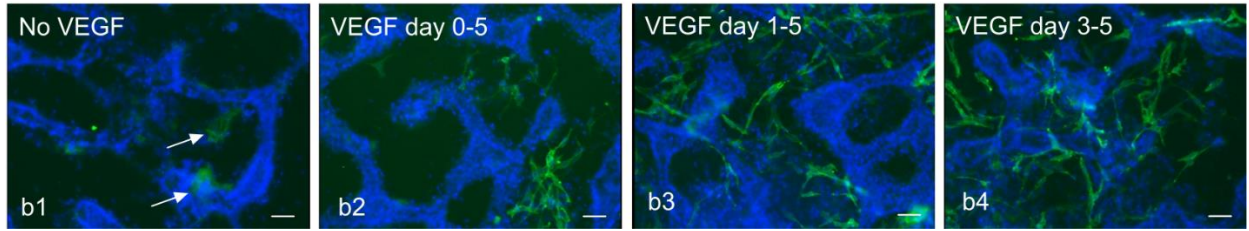

**Figure 2** Treatment conditions to assess the effect of VEGF for different time points. Immunofluorescent staining of the expression of EPC marker CD34 (green) for different time point effect of VEGF. Four treatment conditions were used to assess the different time point effect of VEGF shown in table. The only difference between conditions is the VEGF that was added in different time points. Very low expression (faint staining) of CD34 was observed in the absence of VEGF (Arrows in A). These cells were called precursors of EPCs. Other conditions with different time points of VEGF had clear expression of CD34 which shows the importance of VEGF in induction of EPCs. Interestingly, condition with only 2 days VGEF stimulation (D) had relatively higher CD34 expression compared to 4- and 5-days VEGF stimulation (B & C). n=2 biological replicate and n=6 of technical replicates are shown. The experiment was repeated 3 times with similar results. (n=3 biological replicates) iPSC: induced pluripotent stem cell; ActA: Activin A 25 ng/ml; B4: BMP4, 30 ng/ml; V: (VEGF) 50 ng/ml; BIO: 6-Bromoindirubin-3'-oxime 0.15 μM; S: SB431542 2 μM. Nuclei are stained with DAPI in blue. Scale bar, 64 μm.

To sum up, the final optimisation illustrated that administration of BMP4, Activin A and BIO (Wnt signalling activator) at an early phase induce the cells through the mesoderm lineage. Further stimulation of precursors of PECs (PPECs) (Error! Reference source not found. **c1**) with VEGF and SB431542 (TGFβ-receptor type one inhibitor) during the second phase of treatment leads to efficient differentiation of PECs from human iPSCs within five days (Error! Reference source not found. **c2**). Immunocytochemistry analysis of PECs in this stage illustrated that CD34 and VE-Cadherin expression was inversely related (Error! Reference source not found. **d1-d3**). This suggested that CD34 expressing cells generated with our protocol were more at immature stage of progenitor cells and has not completely differentiated to ECs and were indeed PECs. Furthermore, CD31 and CD34 appeared to be co-localised suggesting that they were vascular endothelial cells (Error! Reference source not found. **d4-d6**). PECs were purified with CD34-labeled beads using MACS technique at day 5 of differentiation. CD34+ cells were plated on Fibronectin-coated plates in EGM-2+ 25% serum containing VEGF 165 (50 ng/ml) and SB 431542 (2 µM) for extra 4 days. Bright Field images of isolated cells after for 4 days in culture shows spindle-like cell morphology, indicated a mature endothelial cell phenotype (Error! Reference source not found.**b** day 10). Furthermore, immunocytochemistry of these cells illustrated that the vast majority of the isolated cells express the endothelial cell marker VE-Cadherin, whereas CD34 expression has been strongly reduced (Error! Reference source not found. **c3 & c3'**). This confirmed endothelial nature of isolated cells and further differentiation of them towards endothelial cells.

It is important to mention that using two iPS batches from the same donor, we found strongly CD34 expressing cells appeared in condition containing only SB 431542 and no VEGF. Further observations revealed that against similar cell number seeded in the first place the cell density in the second set of experiments was much higher after five-day treatment compared to first set of iPS cells. Therefore, it was assumed that because of high confluency and a lack of oxygen, cells might have generated their own VEGF at this stage to compensate for their environmental conditions. To test for this, I used aflibercept (Eylea), vascular endothelial growth factor (VEGF) inhibitor (1:100 dilutions) which is an anti-VEGF drug and cells were treated based on the protocol established earlier. At day three of differentiation, Elyea was added in combination with SB 431542 and was maintained until day five. Then cells were fixed at day five for further immunostaining analysis. Immunostaining results illustrated that in presence of Eylea, the expression of CD34 were highly suppressed (very dim expression) (**Supplementary. Fig**. **7**). Using Eylea resulted in the generation of a homologous population of cells with no indication of CD34 expression. Thus, for our further RNA-sequencing experiments, we used Eylea (VEGF-blocker) to suppress all CD34 expression between day three to five to have more homologous population for non-expressing CD34 positive cells and compared this population with iPS-derived CD34+ cells exposed to VEGF.

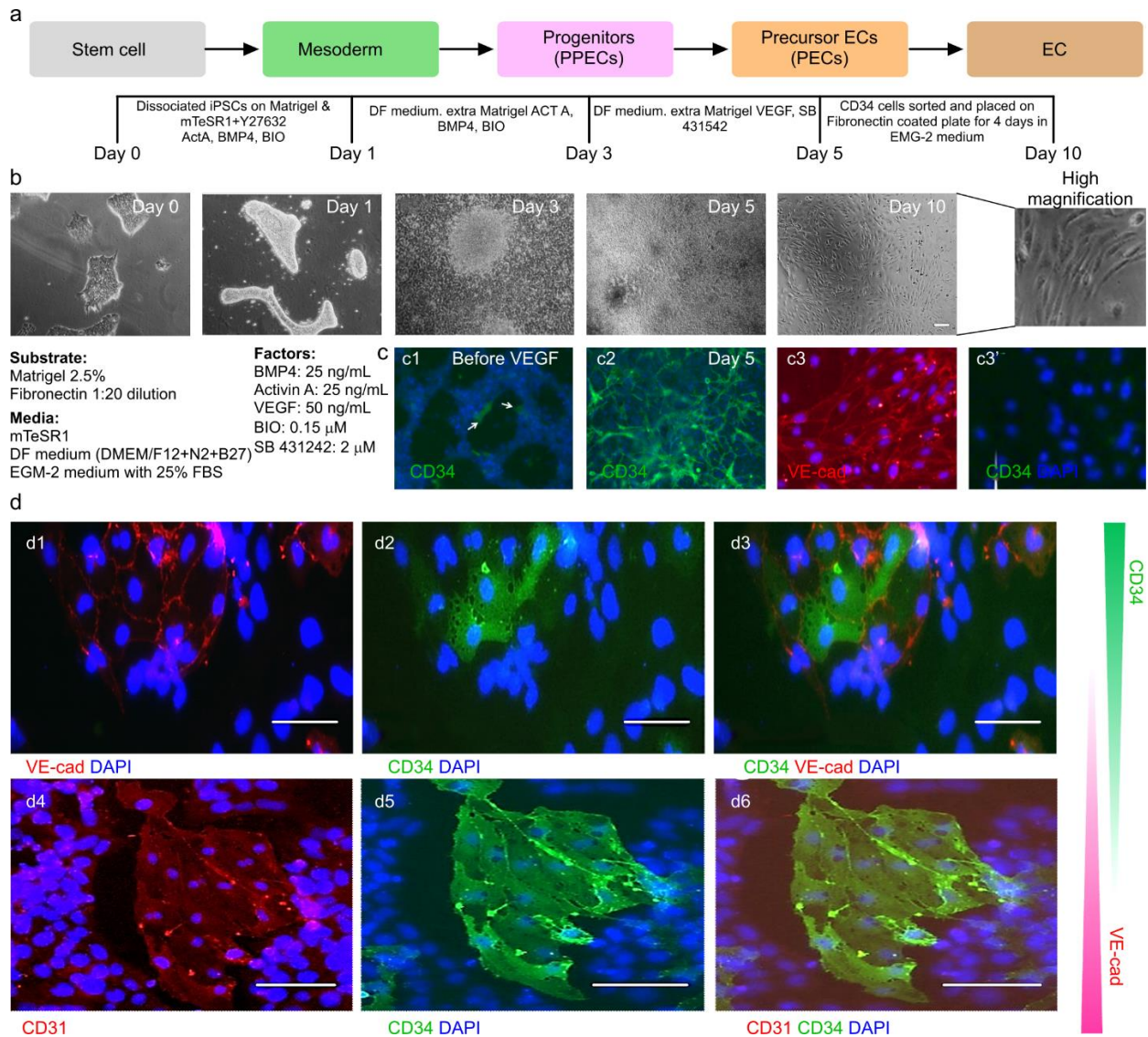

**Figure 3** a) Schematic workflow of PEC differentiation from iPSC cells. b) Bright-field microscopy shows the morphology of the cells at different stages. Undifferentiated iPSC cells in mTeSR1 before treatment at day zero (A), after one day of treatment (B) and after three and five days of treatment, respectively (B & C). Scale bar, 164 μm. c) Representative immunofluorescence pictures of iPSC-derived PECs after 4 days plating on fibronectin. iPSC-derived PECs were purified with CD34-labeled beads at day 5 of differentiation and plated on Fibronectin-coated plates for 4 days. First row, VE-cadherin (red); CD34 (green). Second row, CD31/PE-CAM (red), CD34 (green). Nucleus (DAPI, blue). Experiments were performed two times and in triplicate. CD34 positive cells were negative for VE-Cadherin (A, B and C) which suggests that CD34 expressing cells are more at immature stage of progenitor cells and has not completely differentiated to ECs. Therefore, it could prove that cells are differentiating through the endothelial lineage. Furthermore, CD31 is co-expressed with CD34 at this stage (E, F, and G). Scale bar, 50 μm.

*Transcriptomic comparison of iPSC-derived EPC versus CB and PB*

In order to establish how CD34+ populations from different sources were related to each other, we applied genome wide transcriptional profile. To this end, Truseq illumine RNA sequencing platform was used. For the adult EPC population in this study, we used commercially obtained CD34+ cells that have been purified from adult peripheral blood and from cord blood. Since previous studies illustrated that macrophages/monocytes share lineage traits with EPCs[34] and also cells isolated from macrophages/monocytes could involve in blood vessel regeneration by secreting angiogenic growth factors [35] we also included a sample of CD14+ monocytes to compare its genes transcriptional profiling with the rest of the cell populations. For the iPSC-derived EPCs population we used our protocol to differentiate iPSC towards the EC lineage in the presence of VEGF or a VEGF-blocker. FACS was then used to purify the CD34+ cells from the cultures that contained VEGF. In the case of the cultures that contained the VEGF-blocker, all cells were used without purification. RNA was then isolated from all eight sample populations. Trizol extracted RNA from all samples was then sent to the sequencing facility to Institute of Child Health at University College London (UCL) be processed for RNA-sequencing. Agilent RNA Integrity Number (RIN) was first used to assess RNA quality for RNA sequencing analysis. representative RIN is shown in (**Supplementary Fig**. **7**). After quality control, RNA samples were further proceeded for library preparation and sequencing on the Illumina TruSeq RNA v2 platform. RNAseq data were further normalised by one of our collaborators with bioinformatics expertise (Monte Radeke, UCSB, Santa Barbara, USA). Our RNAseq data was normalised using the TMM method that was applied in the edgeR Bioconductor package (version 2.4.0). The number of counts from RNAseq data were adjusted to reads per million gene alignments per 1kb transcript length (CPMK) to facilitate transparent comparison of transcript levels both within and between samples.

*Validation of RNAseq data*

Once the RNAseq raw data was processed and turned into an excel spreadsheet with 13 columns (for each sample) and around 16000 lines (for different genes), we first checked the different sample populations express specific genes we expected them to express. since CD14 expression was the selection criteria for CD14+ monocytes, we anticipated that gene to be highly expressed in this population. The same applies to CD34 in CD34+ cell populations. As shown in CD14 was strongly expressed in CD14 monocytes while its expression was very low in the rest of the cell populations. Furthermore, CD34 expression was high in all CD34+ populations (Error! Reference source not found. **a)** lined with green and low in CD34- samples, confirming the efficiency of the cell sorting used to purify these samples. The monocyte population is anticipated to strongly express monocyte markers such as CD11b (IT6AM), CD68, LYZ, TYPOBP, FGR, ITGAM, CD4, CD48 and CD36 (Error! Reference source not found. **a)** lined with red. These genes were highly expressed in the CD14 population, which was another confirmation for the validity of the RNAseq data. Furthermore, classic markers of the myeloid lineage (MPO, CD38 and KCNK17) were highly expressed in blood derived CD34+ cells lined in blue (Error! Reference source not found. **a)Error! Reference source not found.**.

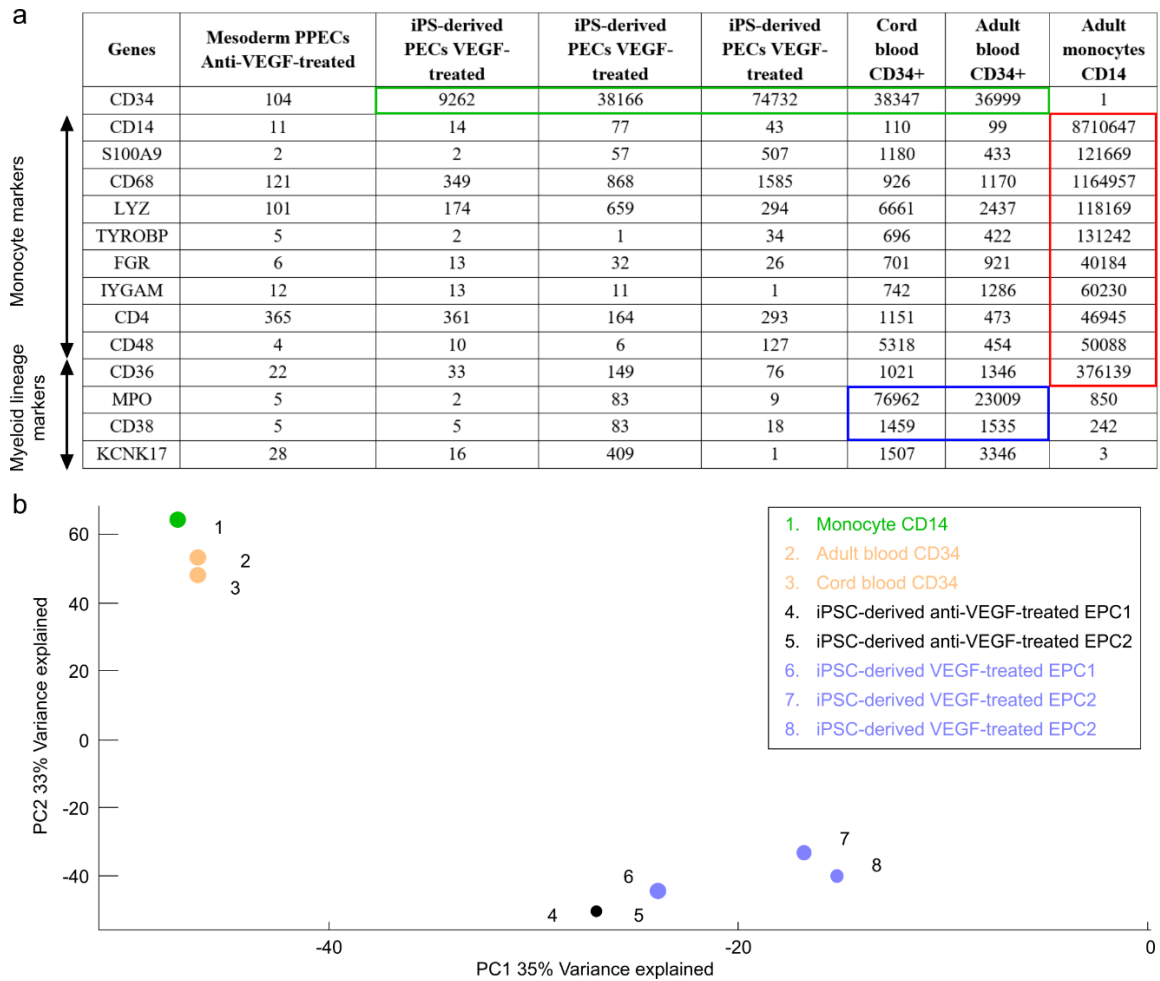

**Figure 4** a) Confirmation of selection criteria by choosing two important genes. CD34 & CD14 were expected to be high in all CD34+ cell population and in CD14 monocytes respectively. Representative genes that are strongly expressed in CD14 monocytes and not in the rest of the cell populations. Blood-derived CD34+ cells are highly expressing classic myeloid markers (MPO, CD38 and KCNK17) compared to the rest of the cell populations. b) Gene expression heat map showing that the two largest gene expression clusters. (Low expression is shown in green, high expression is shown in red). The first largest clusters belong to blood-derived cells and the second group is iPSC-derived PECs, iPSC+ anti-VEGF respectively. c) PCA results show the clustering of each cell population and separation of the clusters according to their gene profiling. PCA analysis demonstrated iPSC-derived EPCs and blood-derived CD34+ cells are fundamentally different cell populations.

After the validation of RNAseq data by choosing selection criteria, genes were further investigated within the groups to get more insight into each specific group. AutoSOME used to create clusters according to the similarity in expression between different groups. Outcome results illustrated that the three blood derived samples including cord blood CD34+, adult blood CD34+ and CD14 monocytes formed the largest cluster (Error! Reference source not found.**b**). The next biggest cluster contained highly expressed genes in iPS-derived CD34+ (VEGF-treated) and iPSC-derived PPECs. These data confirmed that cell populations from the same origin tend to have similar level of gene expressions. For further analysis of these clusters, DAVID online tools were used to find whether there were any biological processes enriched within each cluster[36]. David showed the involvement of the first gene cluster in the regulation of immune response, leukocyte and lymphocytes activation. This is not surprising since these cells are all part of the immune system. However, analysing the other clusters showed their involvement in very different biological pathways. Because AutoSOME[37], did not show how closely related or how far apart the different cell populations are from each other, Principle Component Analysis (PCA) were used in the next step[38]. PCA was performed by using an "R" software module. Then, a plot was generated to visualize the specific gene signatures that represent the similarity between the sample populations. The 8 samples were clustered roughly in 2 different groups. All blood derived cells (cord/adult CD34+ and CD14+ cells) were clustered together (labelled 1, 2 &3) which were clearly segregated from iPS-derived CD34+ and CD34- cells (labelled 4, 5, 6, 7 &8) (Error! Reference source not found. **c**). This data surprisingly demonstrates that adult CD34+ cells are fundamentally different cell population from embryonic derived CD34+ cells. This data is also consistent with AutoSOME data which clustered all blood derived cells together and separated from iPS derived cells. This data suggests that origin of derived cells (embryonic versus adult) is more important than expression of specific marker to indicate the similarities between different cell populations.

**Discussion**

In this study, the main aim was to conduct a gene expression analysis to better understand the nature of EPCs from different embryonic and adult origin. In order to achieve this CD34 was used as marker for EPCs. This marker has been found both in adult circulating EPCs and embryonic EPCs [39-41]. However, because only a small population of EPCs was generated from our stem cells, we first needed to develop an efficient protocol to derive EPCs in larger quantities. Therefore, in the first part, we optimised a protocol to reproducibly differentiate iPS cells into EPCs, which enabled us to isolate these cells in sufficient quantities. This study showed that initial mesodermal commitment was achieved using a cocktail of BMP4, Activin A and BIO in serum-free culture conditions for three days and further EC lineage commitment was achieved with VEGF and SB431542 (Alk5 inhibitor) for two days. Previous studies have shown the importance of BIO (WNT signalling activator) in the presence of Activin A (as a part of TGF-ß) in a very early phase of mesodermal induction [42-44]. The current research also found the presence of these two factors to be crucial in the early phase. Although, BMP4 has been shown to play a significant role in differentiation of stem cells towards a mesodermal lineage [29,45] in our hands administration of BIO and Activin A with or without BMP4 resulted in high expression of CD34. This suggests that BMP4 is not mandatory for the early phase of mesodermal induction. However, the cluster-like morphology of CD34 expressing cells in the presence of BMP4 (versus singly dispersed cells in the absence of BMP4) BMP4 might influence migration and proliferation of already committed precursors rather than their determination.

In contrast, the crucial role of VEGF in endothelial lineage commitment is well-known [46,47]. However, our finding that VEGF is not needed during the early phase of differentiation is less well known. In fact, most investigations aiming to derive ECs or PECs from stem cells add VEGF right from the beginning of their differentiation protocols. Although some previous studies suggested that VEGF likely regulates the survival or propagation of PECs, and not necessarily their differentiation[25,26], our studies could clearly show that VEGF plays a key role in the transition of precursors to PECs. Generated PECs exhibited strong expression of CD34+ cells after 5 days of differentiation treatment. CD34 and VE-Cadherin expression were inversely related at this stage, which indicates our CD34+ cells are progenitors of ECs. Additionally, further culture of purified PECs on fibronectin, differentiated them into completely mature ECs after 4 days. These cells formed a homologous monolayer with cobblestone appearance that exhibited strong expression of VE-Cadherin and confirmed the endothelial nature of our isolated cells.

Another major finding in our study was the discovery that adding extra Matrigel on top of the cells could dramatically enhance the numbers of CD34+ cells in our cultures. This might be caused by an improvement of the 3D matrix around cells, mimicking a more natural microenvironment[48].

Overall, this protocol was shown to be a very efficient method, which was time effective, involved fewer steps compared to other published protocols, required less cell manipulation and was reproducible over repeated experiments. It was demonstrated that at the end of this protocol, over 40% of cells were expressing the CD34 marker. Generated PECs exhibited strong expression of CD34 after five days of differentiation. Isolation of the CD34+ PECs (by MACS) clearly demonstrated that these cells are committed to the EC lineage as they expressed several EC markers.

Two iPS cell batches used in this work were generated by two different lab members but from one donor (BJ iPS cell line). However, beside of the similarity of the protocol and the source of somatic cell line (BJ skin fibroblast) that was used to generate the iPS cells, it appeared that iPS cells behaved slightly different from batch to batch in their proliferation rate and further response to specific factors such as "SB 431542 only condition" during the differentiation protocol. This could demonstrate the sensitivity of each iPS batch and indicates that how the confluency of the cells can affect their response to various growth factors which is essential for directing the differentiation process from early mesendoderm via mesoderm towards a more differentiated PECs. The sensitivity of iPS cells in response to stimulus factors was also evident within one cell line but different passage numbers. It was observed that as the passage number was increasing the capacity of iPS cell to differentiate into PECs were reducing and this was noticeably evident after passage number above 22. The limitations of the cell lines at high passage number should be taken into consideration especially in the level of cell therapy products for therapeutic applications.

Finally, Overall, RNAseq data revealing a distinct transcriptome profile in different CD34+ / CD34- populations, suggested that a five-day differentiation protocol profoundly affected iPSC cells, to differentiate them into endothelial lineage which was reflected in their gene expression levels. After the validation of RNAseq data, by choosing selection criteria, genes were further investigated within the groups to get more insight into each specific group. Our investigations showed high expression levels of monocyte/macrophage markers in CD14+ cells which is not surprising as is know these cells are a part of the immune system. However, some of these genes (TYROBP, FCER1G, HLA-DRA, S100A9, and ITGB2), were also shown to be expressed in so-called "early EPCs"[49]. Our investigations displayed the high expression levels of these genes in the CD14+ cell population. Therefore, this data confirms the validity of previous studies regarding the characteristics of early EPCs, and that they share lineage traits with immune cells, specifically macrophages and monocytes

PCA analysis to decompose the overall variability of gene expression data indicated that iPS-derived PECs and iPS mesoderm (with anti-VEGF) had the greatest similarity in their gene expression levels. This should be due to the fact that these samples came from the same cell sources (+/- VEGF for two days). However, comparing VEGF versus anti-VEGF treated populations, suggest that adding VEGF is sufficient to change the gene expression profile that could be completely distinguishable in an overall gene expression analysis. This could also confirm our in vitro experiments which showed high expression of CD34 cells in VEGF-treated population. Furthermore, PCA analysis grouped cord and blood derived CD34+ cells together with a small distance from CD14 monocytes which indicates reasonably high similarity between these population and suggests that all blood derived cells are closely related and coming from same origin. However, it was surprising to find that cord and peripheral blood derived CD34+ cells were clustered far from the iPSCs. This could suggest that blood derived CD34+ cells are not closely related to mesodermal cell types against what was assumed before and therefore, may have some different stemness properties. Revealing of differences in gene expression pattern between embryonic and adult progenitor cells provides a molecular basis for the discrepancies in the efficiency of different clinical trails and highlight the importance of a detailed definition of these cell types used in clinical trials. Better understanding of these cells will reduce the challenges in isolating heterogeneous population of cells and will help to enhance the potential clinical benefits of EPC based therapy.

Further in vivo experiment would be interesting to find out if injecting of iPS-derived PECs into damaged vasculature will be useful to see whether these cells can integrate into vasculature and differentiate to endothelial cells. In summary, our protocol could be used as a platform to develop EPCs into a more reliable therapeutic products and our RNA-seq data facilities the better understanding of the origin of different EPC populations which will facilitate the translation of regenerative approaches in this field.

**Methods**

*Human induced pluripotent cell culture*

BJ iPS cell line were reprogrammed from human fibroblasts by two independent lab members at UCL institute of ophthalmology. iPSCs used between passage number 10 to 25. Cells were grown on Matrigel coated plates (diluted 1:15 in Knockout DMEM and the final concentration of 1:30) in mTeSR1 medium (STEMCELL Technologies). Routine culture maintenance for BJ iPS cells was prepared when cells were over 70% confluent. Cells were washed with PBS and incubated in Cell Dissociation Reagent (Stem cell technologies) for 2 to 4 minutes in room temperature (RT). Cells were checked every minute under the microscope until the edge of the colonies start to curl. Then Reagent was aspirated and 6ml mTeSR1 was added. Cells were scrapped and split 1:3 ratios in total of 6 ml mTeSR1. Media needs to be changed every day with fresh medium.

*Differentiation protocol to generate iPS-derived CD34+ cells*

Confluent iPSCs dissociated by gentle enzymatic treatment 1X TrypLE (Life technology) for 7 minutes at RT. Cells were then diluted with mTeSR1 and centrifuged at 800 g, for 3 minutes at 25 °C. Cells were re-suspended in 5ml mTeSR1 and Rock inhibitor (Y276332 (Calbiochem). Then cells were counted and seeded onto Matrigel-coated 96 plates at $4\times10^4$ cells/well and left in incubator for 48 hrs. Differentiation was induced two days after by replacing mTeSR1 with differentiation media (DF) (DMEM/F12 + B2 + N27) and timed addition of the following growth factors: 25 ng/ml Activin A (Peprotech, 120-14), 30 ng/ml (BMP) 4 (Peprotech, AF-120-05) and 0.15 µM BIO (TOCRIS, 3194) with extra Matrigel 1:80 final ratio. This medium was refreshed the next day with the same factors. Then factors were replaced to 50 ng/mlVEGF165 (Peprotech, 100-20) and 2 µM SB 431542 (TOCRIS, 1614) at day 3 up to day 5. Cells were fixed at day 5 and immunocytochemistry for CD34 was performed to identify the cells of interest.

*Immunocytochemistry*

Cells were washed with PBS and fixed with 4% paraformaldehyde (PFA) for 10-15 minutes in room temperature (RT). Then cells were washed two times with PBS and then incubated with blocking buffer (blocking buffer: 1% BSA, 0.1% Triton, 0.01% tween 20/PBS.) for 30 minutes. after that, primary antibodies were diluted into blocking buffer and were added per well plates. Primary anti-bodies were incubated for 1hour at RT. Then cells were washed three times with PBS (5 minutes each). Then secondary antibodies IgG (Alexa Fluor 488, 594 Goat anti-mouse, Invitrogen) all were diluted 1/200 in blocking buffer and were added to the wells. Incubation time for secondary antibodies was 1 hr. After incubation time, cells were washed once with PBS for 5 minutes. Then 1 µg/ml dilution of Hoechst was added for nuclei staining only for 30 seconds. Cells were then washed for another 5 minutes with PBS and then were examined under the florescent microscope.

*Microscopy*

For all light imaging microscopy, photomicroscope (Zeicc Axiophot) was used which was attached to a CCD camera (ORCA-ER (Hamamatsu). For each staining, pictures (at least 4 pictures from each plate) were taken with different magnification in an attempt to represent most faithfully. For all Fluorescence microscopies we used up right Axioscope and inverted S100 fluorescence microscopies (Carl Zeiss) were used to analysis the immune-staining results.

*Semi-quantitative method to score the expression of CD34*

Regarding the score method using "number sign" conditions with 1 or 2 small clusters or few dispersed cells were scored with 1 number sign (#), whereas conditions with 2 to 5 small clusters and/or a few numbers of dispersed CD34 positive cells covering around 5-10% of the surface were scored with 2 number sign (# #). Conditions having between 5 to 10 clusters and/or quite high number of dispersed cells covering around 15-20% of the surface were scored with 3 number signs (# # #). Conditions having 10 to 20 small and medium clusters with high number of dispersed CD34 expressing cells covering between 20-30 % of the surface were scored with 4 number signs (# # # #). Furthermore, conditions with considerable number of small/medium clusters and high number of dispersed cells covering between 30-40% of surface were scored with 5 number signs (# # # # #). Similar conditions with comparatively more coverage of CD34 expressing cells between 40-50% of surface were scored with 6 number signs (# # # # # #). Conditions having very strong expression of CD34 covering between 50-60% of the surface containing big clusters of CD34 expression and very high number of dispersed cells were scored with 7 number sign (# # # # # # #) whereas comparatively higher percentage coverage between 60-70% were scored with 8 number sign (# # # # # # # #). Relative expression of CD34 was analysed in a semi-quantitative score method. Relative CD34+ cell yield (number sign), cell detachment (low/ medium/ high) and CD34+ coverage (disperse/ aggregated / extensive) were qualitatively assessed for each treatment condition by two independent observers. Since in some conditions, a big region of cells was dead and detached from the surface, cell detachment was also scored in a semi-quantitative way according to three low, medium and high levels of detachment. Furthermore, it was found that CD34 expressing cells in different conditions had different morphologies. Some conditions had more cluster-like structures whereas others were dispersed cells expressing CD34. Therefore, conditions were also categorised either dispersed, aggregated or in the case of having strong expression of both were categorised as extensive. All experiments with no expression of CD34 were shown by minus sign (-).

*Fluorescence-activated cell sorting (FACS)*

MoFlo XDP (Beckman Coulter) was used to sort different unfixed samples of iPSCs. Cells were labelled with fluorescence conjugated antibodies. FACS buffer was composed of PBS (pH 7.2), 1% BSA and 2mM

EDTA. The whole process was done on ice. A single cell suspension derived by adding 5ml TryPle (1×) and incubation at 37 °C for 10 minutes. Cells were then transferred to 15 ml tube and were centrifuged at 1200 for 5 minutes and 20 °C. Pellet was re-suspended in 90 µL FACS buffer plus 10 µL of FcR-blocking reagent (Miltenyi Biotec) and incubated at 4 °C for 10 minutes. Then, cells were centrifuged in 1200 RCF for 5 minutes and then resuspended in 200 µL of FACS buffer and were transferred on ice to sort on FACS machine. Cells were sorted directly into the Trizol.

*Magnetic Cell Sorting (MACS)*

The MACS CD34 MicroBead Kit (Miltenyi Biotec) used to purify the CD34 expressing cells. After making single suspension similar to FACS, cell then was filtered by Pre-separation filters, 30 µm (Miltenyi Biotec) to prevent cell clumps from clogging the magnetic column, then centrifuged and re-suspended in MACS buffer and counted. Then 100ul MicroBeads conjugated to monoclonal mouse anti-human CD34 antibodies (isotype: mouse IgG1) was added and were incubated in 4°C for 30 minutes. Then cells were centrifuged and re-suspended in MACS buffer and proceeded for magnetic separation with MS columns (Miltenyi Biotec). Cells conjugated with CD34 MicroBeads was loaded to the reservoir of the column. The unlabeled cells flow-through the column collected in a 15ml centrifuge tube. Finally, The MS magnetic column was removed from the magnetic field, 500ul of MACS buffer was loaded in the column reservoir and magnetically labelled cells were flushed out the column by pushing the plunger steadfastly into the reservoir of the column.

*Extraction of RNA for CD34+ and CD34- cells*

During FACS, cells were directly sorted into 1ml of Trizol (Sigma-Aldrich) and then were transferred into 1.5 ml Eppendorfs and were rotated in shaker for 10 minutes in RT to allow lysis to start. Then 200 µL Chloroform (VWR, UK) was added to each Eppendorf, mixed completely and were left at RT for 10 minutes. Tubes were then centrifuged at 12000 RCF for 15 minutes at 4 °C. The top aqueous phase including RNA was removed and placed into the new tube. 1ul of RNase-free glycogen (Ambion, UK) and 500 µL of Isopropanol (VWR, UK) were added to precipitate the RNA. Tubes were inverted to mix and incubated at RT for 10 minutes and then were centrifuged at 12000 RCF at 4 °C for 10 minutes. After centrifuge, the supernatant was poured off and 1ml of 70% Ethanol was added, completely mixed with vortex and centrifuged at 4 °C at 7400 RCF for 5 minutes. after the last centrifuge, supernatant was poured off and the pellet was left to airdry. Then pellet was re-suspended in 21µL of RNAssecure (Ambion, UK) and were heated to 50 °C for 10 minutes.

*RNA-sequencing*

In our project, Illumina TruSeq RNA v2 (Wang et al., 2009) RNA sequencing protocol was used. As a starting material, approximately 250 ng of Trizol extracted RNA was sent to the sequencing facility of UCL Institute of Child Health. Before RNA-sequencing, samples were tested for quality using Agilent RNA Integrity Number (RIN) to determine RNA quality for RNA sequencing analysis. After quality control, RNA samples were further proceeded for library preparation and sequencing on the Illumina TruSeq RNA v2 platform.

**Acknowledgement**

**Author contributions**

W.R. and K.A. conceived the study. W.R. conducted the experiments. W.R. and M.N. analyzed the data. W.R. and M.N. wrote the manuscript and participated in paper revisions.

**Competing financial interests**

The authors declare no competing financial interests.